\documentclass[a4paper,12pt]{article}
\usepackage{bm}
\usepackage{amsmath}
\usepackage{amssymb}
\usepackage[dvips]{graphicx}
\usepackage[dvips]{psfrag}

\makeatletter
\@addtoreset{equation}{section}
\renewcommand{\theequation}{\thesection.\@arabic\c@equation}
\makeatother
\makeatletter
\renewcommand\appendix{\par
  \setcounter{section}{0}%
  \setcounter{subsection}{0}%
  \gdef\thesection{\@Alph\c@section }
  \renewcommand{\theequation}
  {\Alph{section}.\arabic{equation}}
}
\makeatother

\def\nn{\nonumber}
\def\lb{\label}
\def\ci{\cite}
\newcommand{\Ref}[1]{(\ref{#1})}
\def\a{\alpha}
\def\b{\beta}
\def\g{\gamma}

\def\e{\epsilon}

\def\th{\theta}
\def\s{\sigma}

\def\vp{\varphi}

\newcommand{\rd}{\mathrm{d}}
\newcommand{\p}{\partial}
\newcommand{\N}{\nabla}

\def\bra{\langle}
\def\ket{\rangle}
\def\l{\left}
\def\r{\right}

\def\f{\frac}

\def\T{\langle T_{\mu\nu} \rangle}
\def\TU{\langle T_{UU} \rangle}
\def\TV{\langle T_{VV} \rangle}
\def\TUV{\langle T_{UV} \rangle}
\def\Tth{\langle T^\theta{}_\theta \rangle}
\def\Tt{\langle T^\mu{}_\mu \rangle}
\def\cw{c_{\rm w}}
\begin{document}

\begin{titlepage}

\vspace*{-15mm}   
\baselineskip 10pt   
\begin{flushright}   
\begin{tabular}{r} 
\end{tabular}   
\end{flushright}   
\baselineskip 24pt   
\vglue 10mm   

\begin{center}
{\Large\bf A Model of Black Hole Evaporation \\and 4D Weyl Anomaly}
\vspace{8mm}   
\baselineskip 18pt   

\renewcommand{\thefootnote}{\fnsymbol{footnote}}

Hikaru~Kawai$^a$\footnote[2]{hkawai@gauge.scphys.kyoto-u.ac.jp} and 
Yuki~Yokokura$^b$\footnote[4]{yuki.yokokura@riken.jp}

\renewcommand{\thefootnote}{\arabic{footnote}}

\vspace{5mm}   

{\it  
 $^a$ Department of Physics, Kyoto University, 
 Kitashirakawa, Kyoto 606-8502, Japan \\
$^b$ iTHES Research Group, RIKEN, Wako, Saitama 351-0198, Japan}

\vspace{10mm}   

\end{center}

\begin{abstract}
We analyze time evolution of a spherically-symmetric collapsing matter 
from a point of view that black holes evaporate by nature. 
We consider conformal matters and 
solve the semi-classical Einstein equation $G_{\mu\nu}=8\pi G \bra T_{\mu\nu} \ket$
by using the 4-dimensional Weyl anomaly with a large $c$ coefficient. 
Here $\bra T_{\mu\nu} \ket$ contains the contribution from both the collapsing matter and Hawking radiation. 
The solution 
indicates that 
the collapsing matter forms a dense object 
and evaporates without horizon or singularity, 
and it has a surface but looks like an ordinary black hole from the outside. 
Any object we recognize as a black hole should be such an object.

\end{abstract}

\baselineskip 18pt   

\end{titlepage}

\newpage

\section{Introduction and the basic idea} 
Black holes are formed by matters and evaporate eventually \ci{Hawking1}. 
This process should be governed by dynamics of a coupled quantum system of matter and gravity. 
It has been believed for a long time that 
taking the back reaction from the evaporation into consideration 
does not change the classical picture of black holes drastically. 
This is because 
evaporation occurs in the time scale $\sim a^3/l_p^2$ as a quantum effect 
while collapse does in the time scale $\sim a$ as a classical effect 
\footnote{See e.g. \ci{Joshi1} for a classical analysis of collapsing matters.}. 
Here $a=2GM$ and $l_p \equiv \sqrt{\hbar G}$. 
However, these two effects become comparable near the black hole. 
Recently, it has been discussed that 
the inclusion of the back reaction plays 
a crucial role in determining the time evolution of a collapsing matter \ci{KMY,KY1,KY2,Ho1,Ho2,Ho3}. 

We first explain our basic idea by considering the following process. 
Suppose that a spherically symmetric black hole with mass $M=\f{a}{2G}$ is evaporating. 
Then, we consider what happens if we add a spherical thin shell to it. 
The important point here is that the shell will never go across ``the horizon" 
because the black hole disappears before the shell reaches ``the horizon".

To see this, 
we assume for simplicity that Hawking radiation goes to infinity without reflection, 
and then describe the spacetime outside the black hole by the outgoing Vaidya metric \ci{Vaidya}: 
\begin{equation}\lb{Vaidya}
ds^2 =-\f{r-a(u)}{r}du^2-2dudr+r^2d\Omega^2,
\end{equation}
where $M(u)=\f{a(u)}{2G}$ is the Bondi mass. 
We assume that $a(u)$ satisfies 
\begin{equation}\lb{da}
\f{da}{du}=-\f{\s}{a^2},
\end{equation} 
where $\s=kNl_p^2$ is the intensity of the Hawking radiation. 
Here 
$N$ is the degrees of freedom of fields in the theory, 
and $k$ is an $O(1)$ constant.

If the shell comes close to $a(u)$, the motion is governed by 
the equation for ingoing radial null geodesics: 
\begin{equation}\lb{r_u}
\f{dr(u)}{du}=-\f{r(u)-a(u)}{2r(u)}
\end{equation}
no matter what mass and angular momentum the particles consisting the shell have \footnote{See Appendix I in \ci{KY2} for a precise derivation}.
Here $r(u)$ is the radial coordinate of the shell. 
This reflects the fact that 
any particle becomes ultra-relativistic near $r\sim a$ and behaves like a massless particle \ci{Landau_C}. 
As we will show soon in the next section, we obtain the solution of \Ref{r_u}: 
\begin{align}\lb{R0}
r(u)&\approx a(u)-2a(u)\f{d a}{d u}(u)+Ce^{-\f{u}{2a(u)}} \nn \\
 &= a(u)+\f{2\s}{a(u)}+Ce^{-\f{u}{2a(u)}}\longrightarrow a(u)+\f{2\s}{a(u)}~.
\end{align}
This means the followings (see Fig.\ref{particle}.): 
\begin{figure}[h]
 \begin{center}
 \includegraphics*[scale=0.17]{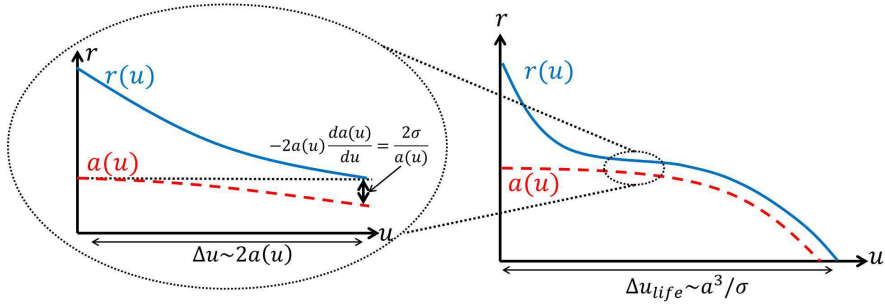}
 \caption{Motion of a shell or a particle near the evaporating black hole.}
 \label{particle}
 \end{center}
 \end{figure}
The shell approaches the radius $a(u)$ in the time scale of $O(2a)$, 
but,  during this time, the radius $a(u)$ itself is slowly shrinking as \Ref{da}. 
Therefore, 
$r(u)$ is always apart from $a(u)$ by $-2a\f{d a}{du}$.
Thus, the shell never crosses the radius $a(u)$
as long as the black hole evaporates in a finite time, 
which keeps the $(u,r)$ coordinates complete outside ``the horizon", $r>a(u)$. 

After the shell comes sufficiently close to $r=a+\f{2\s}{a}$, 
the total system composed of the black hole and the shell 
behaves like an ordinary black hole with mass $M+\Delta M$, 
where $\Delta M$ is the mass of the shell. 
In fact, as we will see later, 
the radiation emitted from the total system agrees with 
that from a black hole with mass $M+\Delta M$. 

\begin{figure}[h]
 \begin{center}
 \includegraphics*[scale=0.22]{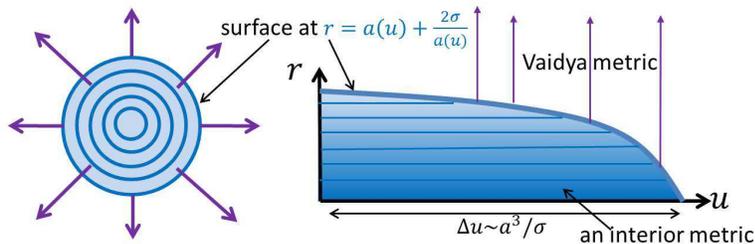}
 \caption{A black hole as an object that consists of many shells.}
 \label{BH}
 \end{center}
 \end{figure}

We then consider a spherically symmetric collapsing matter with a continuous distribution, 
and regard it as a set of concentric null shells. 
We can apply the above argument to each shell 
because its time evolution is not affected by the outside shells due to the spherical symmetry. 
Thus, we conclude that any object we recognize as a black hole actually consists of many shells.
See Fig.\ref{BH}. 
Therefore, there is not a horizon but a surface at $r=a+\f{2\s}{a}$, 
which is a boundary inside which the matter is distributed
\footnote{What is essential for particle creation is 
a time-dependent metric but not the existence of horizons. 
A Planck-like distribution can be obtained even if there is no horizon \ci{KMY,KY2,Barcelo1}.}. 
If we see the system from the outside, it looks like an evaporating black hole 
in the ordinary picture. 
However, it has a well-defined internal structure in the whole region, 
and evaporates like an ordinary object \footnote{We keep using the term ``black hole" 
even though the system is different from the conventional black hole that has a horizon.}
\footnote{See also \ci{Barcelo,Allahbakhshi,Baccetti1,Baccetti2}. 
See e.g. \ci{Frolov,Bambi} for a black hole as a closed trapped region in the vacuum.}.

In order to prove this idea, we have to analyze 
the dynamics of the coupled quantum system of matter and gravity. 
As a first step, we consider the self-consistent equation
\begin{equation}\lb{Einstein}
G_{\mu\nu}=8\pi G \bra T_{\mu\nu} \ket.
\end{equation}
Here we regard matter as quantum fields while we treat gravity as a classical metric $g_{\mu\nu}$. 
$\bra T_{\mu\nu} \ket$ is the expectation value of 
the energy-momentum tensor operator with respect to the state $|\psi \ket $ that 
stands for the time evolution of matter fields defined on the background $g_{\mu\nu}$.
$\bra T_{\mu\nu} \ket$ contains the contribution from both the collapsing matter and the Hawking radiation, 
and $|\psi \ket$ is any state that represents a collapsing matter at $u=-\infty$.

In this paper, we consider conformal matters. 
Then, we show that $\bra T_{\mu\nu} \ket$ on an arbitrary spherically symmetric metric $g_{\mu\nu}$ 
can be determined by the 4-dimensional (4D) Weyl anomaly with some assumption, 
and obtain the self-consistent solution of \Ref{Einstein} 
that realizes the above idea. 
Furthermore, we can justify that the quantum fluctuation of gravity is small 
if the theory has a large $c$ coefficient in the anomaly.

Our strategy to obtain the solution is as follows. 
We start with a rather artificial assumption that $\bra T^t{}_t \ket+\bra T^r{}_r \ket=0$. 
(This is equivalent to $\bra T_{UV} \ket =0$ in Kruskal-like coordinates.)
By a simple model satisfying this assumption, we construct a candidate metric $g_{\mu\nu}$. 
We then evaluate $\bra T_{\mu\nu} \ket$ on this background $g_{\mu\nu}$
by using 
the energy-momentum conservation and the 4D Weyl anomaly, 
and show that the obtained $g_{\mu\nu}$ and $\bra T_{\mu\nu} \ket$ satisfy \Ref{Einstein}. 
Next, we try to remove the assumption. 
We fix the ratio $\bra T^r{}_r \ket/\bra T^t{}_t \ket$, 
which seems reasonable for the conformal matter. 
Under this ansatz, the metric is determined from 
the trace part of \Ref{Einstein}, 
$G^\mu{}_\mu=8\pi G \bra T^\mu{}_\mu \ket$, 
where $\bra T^\mu{}_\mu \ket$ is given by the 4D Weyl anomaly. 
On this metric, we calculate $\bra T_{\mu\nu} \ket$ as before, 
and check that \Ref{Einstein} indeed holds. 

This paper is organized as follows. 
In section \ref{s:Motion} 
we derive \Ref{R0}. 
In section \ref{s:Metric} 
we construct a candidate metric with the assumption $\bra T^t{}_t \ket+\bra T^r{}_r \ket=0$. 
In section \ref{s:EMT} we evaluate $\bra T_{\mu\nu} \ket$ on this metric, 
and then check that \Ref{Einstein} is satisfied. 
In section \ref{s:Gen} we remove the assumption 
and construct the general self-consistent solution. 
In section \ref{s:radiation} we rethink how the Hawking radiation is created in this picture. 

\section{Motion of a thin shell near the evaporating black hole}\lb{s:Motion}
We start with the derivation of \Ref{R0} \ci{KMY,KY1,KY2}. 
That is, we solve \Ref{r_u} explicitly. 
Putting $r(u)=a(u)+\Delta r(u)$ in \Ref{r_u} and assuming $\Delta r(u)\ll a(u)$, 
we have 
\begin{equation}\lb{r_u2}
\f{d \Delta r(u)}{d u} = -\f{\Delta r(u)}{2a(u)}-\f{d a(u)}{d u}~.
\end{equation}
The general solution of this equation is given by 
\begin{equation}
\Delta r(u)=C_0 e^{-\int^u_{u_0}d u' \f{1}{2a(u')}} + \int^u_{u_0}d u' \l(-\f{d a}{d u}(u') \r)e^{-\int^{u}_{u'}d u''\f{1}{2a(u'')}}, \nn
\end{equation}
where $C_0$ is an integration constant. 
Because $a(u)$ and $\f{d a(u)}{d u}$ can be considered to be constant in the time scale of $O(a)$,
the second term can be evaluated as 
\begin{align*}
 &\int^u_{u_0}d u' \l(-\f{d a}{d u}(u') \r)e^{-\int^{u}_{u'}d u''\f{1}{2a(u'')}}\\
 &\approx -\f{d a}{d u}(u) \int^u_{u_0}d u' e^{-\f{u-u'}{2a(u)}}=-2\f{d a}{d u}(u) a(u)(1-e^{-\f{u-u_0}{2a(u)}}). 
\end{align*}
Therefore, we obtain 
\begin{equation}
\Delta r(u)\approx C_0 e^{-\f{u-u_0}{2a(u)}} -2\f{d a}{d u}(u)a(u) (1-e^{-\f{u-u_0}{2a(u)}}), \nn 
\end{equation}
which leads to \Ref{R0}: 
\begin{align}
r(u)&\approx a(u)-2a(u)\f{d a}{d u}(u)+Ce^{-\f{u}{2a(u)}} \nn \\
 &= a(u)+\f{2\s}{a(u)}+Ce^{-\f{u}{2a(u)}}\longrightarrow a(u)+\f{2\s}{a(u)}.\nn
\end{align}
This result indicates that any particle gets close to 
\begin{equation}\lb{R}
R(a)\equiv a +\f{2\s}{a}
\end{equation}
in the time scale of $O(2a)$, 
but it will never cross the radius $a(u)$ as long as $a(u)$ keeps decreasing as \Ref{da}
\footnote{
The above analysis is based on the classical motion of particles, 
but we can show that the result is valid even if we treat them quantum mechanically. 
See section 2-B and appendix A in \ci{KY2}.}. 
In the following we call $R(a)$ the surface of the black hole. 

Here one might wonder if such a small radial difference $\Delta r=\f{2\s}{a}$ makes sense, 
since it looks much smaller than $l_p$. 
However, the proper distance between the surface $R(a)$ and the radius $a$ 
is estimated for the metric \Ref{Vaidya} as
\footnote{For the general metric, 
the proper length in the radial direction is given by 
$\Delta l = \sqrt{g_{rr}-\f{(g_{ur})^2}{g_{uu}}}\Delta r$. See \ci{Landau_C}.}
\begin{equation}\lb{dl}
\Delta l
=\sqrt{\f{R(a)}{R(a)-a}}\f{2\s}{a}\approx\sqrt{2\s}~.
\end{equation}
In general, this is proportional to $l_p$, 
but it can be large if we consider a theory with many species of fields. 
In fact, in that case we have 
\begin{equation}\lb{largeN}
\s\sim N l_p^2\gg l_p^2.
\end{equation}
We assume that $N$ is large but not infinite, for example, 
$O(100)$ as in the standard model. 
Then, $\Delta r=\f{2\s}{a}$ is a non-trivial distance. 

\section{Constructing the candidate metric}\lb{s:Metric}
The purpose of this section is to construct 
a candidate metric by considering a simple model corresponding 
to the process given in section 1 \ci{KMY, KY2}.
At this stage, we don't mind whether it is a solution of \Ref{Einstein} or not, 
which will be the task for the next section. 
\subsection{Single-shell model}
\begin{figure}[h]
 \begin{center}
 \includegraphics*[scale=0.18]{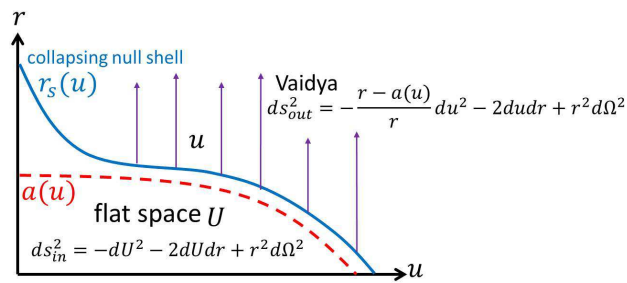}
 \caption{A spherical null shell evaporating in accordance with \Ref{da}.}
 \label{single}
 \end{center}
 \end{figure}

As a preliminary for the next subsection, we begin with a simpler model \ci{KMY}. 
See Fig.\ref{single}.
Suppose that a spherical null shell with mass $M=\f{a}{2G}$ comes from infinity, 
and evaporates like the ordinary black hole. 
Here we consider the shell infinitely thin. 
We model this process by describing the spacetime outside the shell as the Vaidya metric \Ref{Vaidya} with \Ref{da}. 
On the other hand, the spacetime inside it is flat because of spherical symmetry, 
and we express the metric by 
\begin{equation}\lb{flat}
ds^2 = - dU^2 -2dU dr + r^2 \Omega^2. 
\end{equation}

Now we have two time coordinates $(u,U)$, 
and we need to connect them along the trajectory of the shell, $r=r_s(u)$. 
This can be done by noting that
the shell is moving along an ingoing null geodesic in the metrics of the both sides, 
\Ref{Vaidya} and \Ref{flat}. 
Therefore, the junction condition is given by 
\begin{equation}\lb{junction}
\f{r_s(u)-a(u)}{r_s(u)}du=-2dr_s = dU. 
\end{equation}
This determines the relation between $U$ and $u$ for a given $a(u)$. 

Generally, connecting two different metrics along a null hypersurface $\Sigma$ leads to 
a surface energy-momentum tensor $T^{\mu\nu}_{\Sigma}$. 
Indeed, by using the Barrabes-Israel formalism \ci{Israel_null,Poisson}, 
we can estimate the surface energy $\epsilon_{2d}$ and the surface pressure $p_{2d}$ as
\footnote{The surface tensor is given by 
$ T^{\mu\nu}_{\Sigma} = (-\bm k\cdot \bm v)^{-1}\delta (\tau) \left(  \epsilon_{2d} k^{\mu} k^{\nu} + p_{2d} \sigma^{\mu\nu} \right)$.
Here $\bm v=\frac{\partial}{\partial \tau}$ is the 4-vector of a timelike observer 
with proper time $\tau$ who crosses the shell at $\tau =0$, 
$\bm k$ is the ingoing radial null vector along the locus of the shell 
which is taken as $\bm k=\frac{2r_s(u)}{r_s(u)-a(u)}\partial_u-\partial_r$ for $r>r_s$ and 
$\bm k=2\partial_{U}-\partial_r$ for $r<r_s$, 
and $\sigma^{\mu\nu}$ is the metric on the 2-sphere ($\sigma_{\mu\nu}dx^{\mu}dx^{\nu}=r^2d\Omega^2$). 
See Appendix F in \ci{KY2} for the detail.}
\begin{equation}\label{EMT_surface}
\epsilon_{2d} = \frac{a}{8\pi G r_s^2},~~~p_{2d} = \frac{-\dot{a}r_s}{4\pi G(r_s-a)^2}.
\end{equation}
Note that $\epsilon_{2d}$ is nothing but the energy per unit area of the shell with energy $M=\f{a}{2G}$, 
and that the positive pressure $p_{2d}$ is proportional to the energy being lost, $-\dot a(u)>0$. 

Thus, we have obtained the metric without coordinate-singularity 
that describes the formation and evaporation process of a black hole. 
Note again that we don't claim yet that this metric satisfies \Ref{Einstein}, 
but we here construct a candidate metric which formally expresses such a process.

\subsection{Multi-shell model}
\begin{figure}[h]
 \begin{center}
 \includegraphics*[scale=0.18]{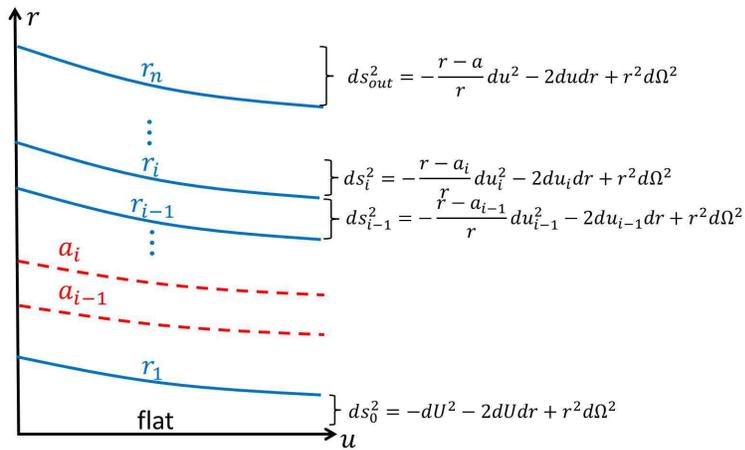}
 \caption{A multi-shell model.}
 \label{multi}
 \end{center}
 \end{figure}

Now, we consider a spherically-symmetric collapsing matter consisting of $n$ spherical thin null shells. 
See Fig.\ref{multi}, where the position of the $i$-th shell is depicted by $r_i$.
We assume that each shell behaves like the ordinary evaporating black hole if we look at it from the outside.
We postulate again that the radiation goes to infinity without reflection.  
Then, because of spherical symmetry, 
the region just outside the $i$-th shell can be described by the Vaidya metric: 
\begin{equation}\lb{Vaidya_i}
ds^2_i =-\f{r-a_i(u_i)}{r}du_i^2-2du_idr+r^2d\Omega^2
\end{equation}
with 
\begin{equation}\lb{da_i}
\f{da_i}{du_i}=-\f{\s}{a_i^2}
\end{equation}
for $i=1\cdots n$. 
Here, $a_i=2Gm_i\gg l_p$, and 
$m_i$ is the energy inside the $i$-th shell (including the contribution from the shell itself). 
For $i=n$, $u_n=u$ is the time coordinate at infinity, 
and $a_n=a=2GM$, where $M$ is the Bondi mass for the whole system. 
On the other hand, the center, which is below the $1$-st shell, is the flat spacetime \Ref{flat}:
\begin{equation}\lb{flat2}
a_0=0,~~~u_0=U.
\end{equation}

In this case, the junction condition \Ref{junction} is generalized to 
\begin{equation}\lb{junction_i}
\f{r_i-a_i}{r_i}du_i =-2dr_i =\f{r_i-a_{i-1}}{r_i}du_{i-1}~~~{\rm for}~i=1\cdots n.
\end{equation}
This is equivalent to 
\begin{equation}\lb{r_u_i}
\f{dr_i(u_i)}{du_i}=-\f{r_i(u_i)-a_i(u_i)}{2r_i(u_i)} 
\end{equation}
and 
\begin{equation}\lb{u_i}
\f{\rd u_i}{\rd u_{i-1}}=\f{r_i-a_{i-1}}{r_i-a_i}=1+\f{a_i-a_{i-1}}{r_i-a_i}.
\end{equation}

As in the single-shell model, 
we have the surface energy-momentum tensor on each shell. 
By generalizing \Ref{EMT_surface}, 
we can show that the energy density $\epsilon_{2d}^{(i)}$ and the surface pressure $p_{2d}^{(i)}$ 
on the $i$-th shell are given by \ci{KY2}
\begin{equation}\lb{EMT_s_i}
 \epsilon_{2d}^{(i)} = \frac{a_i-a_{i-1}}{8\pi G r_i^2},~~~
 p_{2d}^{(i)} = -\frac{r_i}{4\pi G(r_i-a_i)^2}\l(\f{da_i}{du_i}-\l(\f{r_i-a_i}{r_i-a_{i-1}}\r)^2\f{da_{i-1}}{du_{i-1}}\r).
\end{equation}
$\e_{2d}^{(i)}$ expresses the energy density of the shell with energy $m_i = \f{a_i-a_{i-1}}{2G}$. 
In the expression of $p_{2d}^{(i)}$, 
the first term corresponds to the total energy flux observed just above the shell, 
and the second one represents the energy flux below the shell
that is redshifted due to the shell. 
Thus, the pressure is induced by the radiation from the shell itself
\footnote{See \ci{KY2} for more detailed discussions.}.

\subsection{The candidate metric}
Finally, we take the continuum limit in the multi-shell model and construct the candidate metric \ci{KMY,KY1,KY2}. 
Especially, we focus on a configuration in which 
each shell has already come close to $R(a_i)$: 
\begin{equation}\lb{R_i}
r_i=R(a_i)=a_i+\f{2\s}{a_i},
\end{equation}
where \Ref{R} has been used
\footnote{Due to the spherical symmetry, 
the motion of each shell in the ``local time" $u_i$ is determined independently of the shells outside it. 
Therefore, the analysis for \Ref{R} can be applied to each shell.}. 
(A more general case is discussed in \ci{Ho3}.)

We first solve the equations \Ref{junction_i}.
By introducing 
\begin{equation}\lb{eta_def}
\eta_i\equiv \log \f{\rd U}{\rd u_i},
\end{equation}
we have 
\begin{align}
\eta_i-\eta_{i-1} &=\log \f{ \f{\rd U}{\rd u_i} }{ \f{\rd U}{\rd u_{i-1}}} =- \log \f{\rd u_i}{\rd u_{i-1}}\nn\\
 &=-\log \l( 1+\f{a_i-a_{i-1}}{r_i-a_i} \r)  \nn\\
 &\approx -\f{a_i-a_{i-1}}{r_i-a_i} 
 =-\f{a_i-a_{i-1}}{\f{2\s}{a_i}} \nn\\
 &\approx - \f{1}{4\s} \l(a_i^2-a_{i-1}^2\r)~.
\end{align}
Here, at the second line, we have used \Ref{u_i}; 
at the third line, we have used \Ref{R_i} and assumed $\f{a_i-a_{i-1}}{\f{2\s}{a_i}}\ll 1$, 
which is satisfied for a continuous distribution; 
and at the last line, we have approximated $2a_i\approx a_i+a_{i-1}$. 
With the initial conditions \Ref{flat2}, we obtain
\begin{equation}\lb{eta}
\eta_i =-\f{1}{4\s}a_i^2.
\end{equation}

Now, the metric at a spacetime point $(U,r)$ inside the object 
is obtained by considering the shell that passes the point  
and evaluating the metric \Ref{Vaidya_i}. 
We have at $r=r_i$ 
\begin{align}\lb{evalu1}
\f{r-a_i}{r} &= \f{r_i-a_i}{r_i}=\f{\f{2\s}{a_i}}{r_i}\approx \f{2\s}{r^2} \\
\lb{evalu2}
\f{du_i}{dU} &=e^{-\eta_i}=e^{\f{a_i^2}{4\s}}\approx e^{\f{r^2}{4\s}}, 
\end{align}
where \Ref{R_i} and \Ref{eta} have been used. 
From these, we obtain the metric 
\begin{align}\lb{Ur}
ds^2 &=-\f{r-a_i}{r}du_i^2-2du_idr+r^2d\Omega^2  \nn\\
 &= -\f{r_i-a_i}{r_i}\l(\f{du_i}{dU} \r)^2dU^2-2\l(\f{du_i}{dU} \r)dUdr+r^2d\Omega^2\nonumber \\
 &\approx - \f{2\s}{r^2}e^{\f{r^2}{2\s}} dU^2 -2 e^{\f{r^2}{4\s}}dUdr +r^2 d\Omega^2. 
\end{align}
Note that this is static although each shell is shrinking, 
and that it does not exist in the classical limit $\s \rightarrow 0$. 

Thus, our candidate metric for the evaporating black hole is given by 
\begin{equation}\lb{eva_metric}
d s^2=\begin{cases}
-\f{2\s}{r^2} e^{- \f{R(a(u))^2-r^2}{2\s}} du^2 -2 e^{- \f{R(a(u))^2-r^2}{4\s}} d u dr 
 + r^2 d \Omega^2,~~{\rm for}~~r\leq R(a(u))~,\\
 - \f{r-a(u)}{r}d u^2 -2d r d u + r^2 d \Omega^2,~~{\rm for}~~r\geq R(a(u)),
\end{cases}
\end{equation}
which corresponds to Fig.\ref{BH}.
Here we have converted $U$ to $u$ by $du=e^{\f{R(a(u))^2}{4\s}}dU$ and expressed \Ref{Ur} in terms of $u$. 
This metric is continuous at the surface $r=R(a(u))=a(u) + \f{2\s}{a(u)}$, 
where $a(u)$ decreases as \Ref{da}. 

Next we consider a stationary black hole. 
Suppose that we put this object into the heat bath with temperature $T_H=\f{\hbar}{4\pi a}$. 
Then, the ingoing energy flow from the bath and 
the outgoing one from the object become balanced each other
\footnote{We can see how this ``equilibration" occurs, 
by introducing interactions between radiations and matters. 
See section 2-E in \ci{KY2} for a detailed discussion.}, 
and the system reaches a stationary state, 
which corresponds to a stationary black hole in the heat bath \ci{GH}. 
(See also Fig.\ref{boundary}.)
The object has its surface at $r=R(a)$, where $a=$const.
Then, the Vaidya metric for the outside spacetime is replaced with the Schwarzschild metric: 
\begin{equation}\lb{Sch}
ds^2 =- \f{r-a}{r}d t^2 + \f{r}{r-a}d r^2 + r^2 d \Omega^2. 
\end{equation}
By introducing the time coordinate $T$ around the origin as 
\begin{equation}\lb{T}
dT=dU+\f{r^2}{2\s}e^{-\f{r^2}{4\s}}dr,
\end{equation}
we can write the interior metric \Ref{Ur} as 
\begin{equation}\lb{Tr}
ds^2= - \f{2\s}{r^2}e^{\f{r^2}{2\s}} dT^2 + \f{r^2}{2\s} dr^2 +r^2 d\Omega^2. 
\end{equation}
Thus, by changing $T$ to $t$ through $dt=e^{\f{R(a)^2}{4\s}}dT$, 
we obtain our candidate metric for the stationary black hole:
\begin{equation}\lb{sta_metric}
d s^2=\begin{cases}
-\f{2\s}{r^2} e^{- \f{R(a)^2-r^2}{2\s}} d t^2 + \f{r^2}{2\s} d r^2 + r^2 d \Omega^2,~~{\rm for}~~r\leq R(a)~,\\
 - \f{r-a}{r}d t^2 + \f{r}{r-a}d r^2 + r^2 d \Omega^2,~~{\rm for}~~r\geq R(a)~,
\end{cases}
\end{equation}
where $R(a)=a + \f{2\s}{a}$ with $a=$const.
The remarkable feature of \Ref{sta_metric} is that 
the redshift is exponentially large inside and 
time is almost frozen in the region deeper than the surface by $\Delta r\gtrsim \f{\s}{a}$. 

\section{Evaluating the expectation value of the energy-momentum tensor}\lb{s:EMT}
In this section we evaluate the expectation value of the energy-momentum tensor $\bra T_{\mu\nu} \ket$ 
in the candidate metrics \Ref{eva_metric} and \Ref{sta_metric} 
assuming that the matter is conformal. 
We show that $\bra T_{\mu\nu} \ket$ can be determined by the 4-dimensional Weyl anomaly 
and the energy-momentum conservation $\N^\mu \bra T_{\mu\nu} \ket=0$ 
if we introduce a rather artificial assumption $\bra T_{UV} \ket =0$. 
Then, we show that the self-consist equation \Ref{Einstein} is indeed satisfied 
if $\s$ in \Ref{eva_metric} and \Ref{sta_metric} is chosen properly. 

\subsection{Summary of the assumptions so far}
We start with summarizing the assumptions which we have made to obtain the metric \Ref{eva_metric}. 
Firstly, we assume that the system is spherically symmetric. 
Then, the time evolution of each shell is not affected by its exterior region 
after it becomes ultra-relativistic. 
Secondly, we assume that the radiation coming out of each shell flows to infinity without reflection. 
Then, the metric of each inter-shell region is given by the Vaidya metric. 

We consider what these assumptions mean in terms of $\bra T_{\mu\nu} \ket$. 
Here we discuss in Kruskal-like coordinates $(U,V)$:  
$U$ and $V$ are coordinates such that 
outgoing and ingoing null lines are characterized by $U=$const. and $V=$const., respectively. 
Therefore, the second assumption means that 
in the inter-shell regions only $\bra T_{UU} \ket$ is nonzero
\footnote{We can see this explicitly as follows. 
Because the Vaidya metric has only $G_{uu}$, 
we can expect that only $\bra T_{uu} \ket$ exists in the inter-shell regions. 
From the definitions of $U$ and $V$, 
we have a transformation between $(u,r)$ and $(U,V)$ such that 
$\l (\f{\p u}{\p V}\r)_{U}=0.$
Therefore, we evaluate 
$\bra T_{UU} \ket=\l(\f{\p u}{\p U} \r)^2\bra T_{uu} \ket \neq0$, 
$\bra T_{UV} \ket=\l(\f{\p u}{\p U} \r)\l(\f{\p u}{\p V} \r)\bra T_{uu} \ket=0$
and $\bra T_{VV} \ket=\l(\f{\p u}{\p V} \r)^2\bra T_{uu} \ket =0$.}, 
and in particular, 
\begin{equation}\lb{T_UV}
\bra T_{UV} \ket=0.
\end{equation}
Furthermore, noting the surface energy-momentum tensor \Ref{EMT_s_i},
we find that $\epsilon_{2d}^{(i)}$ and $p_{2d}^{(i)}$ lead to nonzero values of 
$\bra T_{VV} \ket$ and $\bra T^\th{}_\th \ket=\bra T^\phi{}_\phi \ket$, respectively, on each shell.  
(See the footnote at \Ref{EMT_surface}.)

Thus, after taking the continuum limit, we have nonzero values for $\bra T_{\mu\nu} \ket$
except for $\bra T_{UV} \ket$.
Therefore, the assumption we have made so far are 
essentially the spherical symmetry and \Ref{T_UV}. 
We keep the assumption \Ref{T_UV} within this section, and 
will remove it in the next section. 

\subsection{Relations among $\T$ from the energy-momentum conservation} 
We investigate the relations among the components of $\T$ 
obtained from the energy-momentum conservation, 
which will be used to determine $\T$. 
The general spherically symmetric metric can be expressed in Kruskal-like coordinates as 
\begin{equation}\lb{UV}
ds^2 =-e^{\vp(U,V)}dUdV+r(U,V)^2d\Omega^2.  
\end{equation}
We assume that $\bra T_{\mu\nu} \ket$ is spherically symmetric, that is, 
the non-zero components are 
\begin{equation}\lb{assume_T1}
\bra T_{UU} \ket,~~\bra T_{VV} \ket,~~\bra T_{UV} \ket,~~\bra T^\th{}_\th \ket=\bra T^\phi{}_\phi \ket, 
\end{equation}
which depend only on $U$ and $V$. 
Here we keep $\bra T_{UV} \ket$ for the convenience of the next section. 
Then, $\N^\mu \bra T_{\mu U} \ket=0$ and $\N^\mu \bra T_{\mu V} \ket=0$ are expressed as, respectively, 
\begin{equation}\lb{con_U}
\bra T^\th{}_\th \ket=-\f{e^{-\vp}}{r\p_Ur}\l[ \p_V(r^2\bra T_{UU}\ket)+\p_U(r^2 \bra T_{UV}\ket)-\p_U\vp (r^2 \bra T_{UV}\ket) \r],
\end{equation}
\begin{equation}\lb{con_V}
\bra T^\th{}_\th \ket=-\f{e^{-\vp}}{r\p_Vr}\l[ \p_U(r^2\bra T_{VV}\ket)+\p_V(r^2 \bra T_{UV}\ket)-\p_V\vp (r^2 \bra T_{UV}\ket) \r].
\end{equation}
The other components are satisfied trivially. 

On the other hand, because the trace of the energy-momentum tensor is expressed as
$\bra T^\mu{}_\mu \ket =2g^{UV}\bra T_{UV} \ket + 2\bra T^\th{}_\th \ket$, 
we have 
\begin{equation}\lb{trace2}
\bra T^\th{}_\th \ket = \f{1}{2} \bra T^\mu{}_\mu \ket + 2e^{-\vp} \bra T_{UV} \ket.
\end{equation}
Substituting \Ref{trace2} into \Ref{con_U} and \Ref{con_V}, we obtain
\begin{equation}\lb{con_Ut}
\p_U(r^2 \TUV)- \l(\p_U\vp - \f{2}{r}\p_U r \r)(r^2 \TUV) = - \p_V (r^2\TU)- \f{1}{2} r\p_U r e^\vp \Tt,
\end{equation}
\begin{equation}\lb{con_Vt}
\p_V(r^2 \TUV)- \l(\p_V\vp - \f{2}{r}\p_V r \r)(r^2 \TUV) = - \p_U (r^2\TV)- \f{1}{2} r\p_V r e^\vp \Tt.
\end{equation}
Once $\Tt$ is given, 
we can determine $\T$ from these equations with some boundary conditions  
if one of the four functions \Ref{assume_T1} is known \ci{Christensen}. 

\subsubsection{The static case}
As a special case, we suppose that the spacetime is static. 
Then, $\vp(U,V)$ and $r(U,V)$ satisfy 
\begin{equation}\lb{static}
\vp(U,V)=\vp(r(U,V)),~~~\p_V r=-\p_Ur. 
\end{equation}
Then, we can rewrite \Ref{UV} as
\begin{equation}\lb{Tr_ansazt}
ds^2 = - \f{1}{B(r)}e^{A(r)}dT^2 + B(r)dr^2 + r^2 d\Omega^2,
\end{equation}
where
\begin{equation}\lb{UV2}
e^{\vp(r)}= \f{e^{A(r)}}{B(r)},~~~\p_V r=-\p_Ur=\f{e^{\f{A(r)}{2}}}{2B(r)}
\end{equation}
and 
\begin{equation}\lb{UV_T}
dU=dT-Be^{-\f{A}{2}}dr,~~~dV=dT+Be^{-\f{A}{2}}dr. 
\end{equation}

In this case, the expectation value of the energy-momentum tensor $\T$ should also be static
and satisfy 
\begin{equation}\lb{T_static}
\bra T_{\mu\nu} \ket=\bra T_{\mu\nu}(r) \ket,~~\TU=\TV.
\end{equation}
Then, the formulae \Ref{con_Ut} and \Ref{con_Vt} reduce to 
\begin{equation}\lb{formula}
\p_r (r^2\bra T_{UV}\ket) - \l(\p_r \vp - \f{2}{r} \r) (r^2\bra T_{UV}\ket)
= \p_r (r^2\bra T_{UU}\ket) -\f{1}{2}r e^{\vp} \bra T^\mu{}_\mu \ket.
\end{equation}

\subsection{Evaluation of $\bra T_{\mu\nu} \ket$ inside the black hole}
Now we can evaluate $\T$ in the metric \Ref{Tr} assuming \Ref{T_UV} and \Ref{T_static}. 
Here we rewrite the metric \Ref{Tr} as \Ref{UV} with \Ref{UV2} and
\begin{equation}\lb{AB_sta}
A(r)=B(r)=\f{r^2}{2\s}. 
\end{equation}

\subsubsection{Boundary conditions for $\T$}
\begin{figure}[h]
\begin{center}
\includegraphics*[scale=0.23]{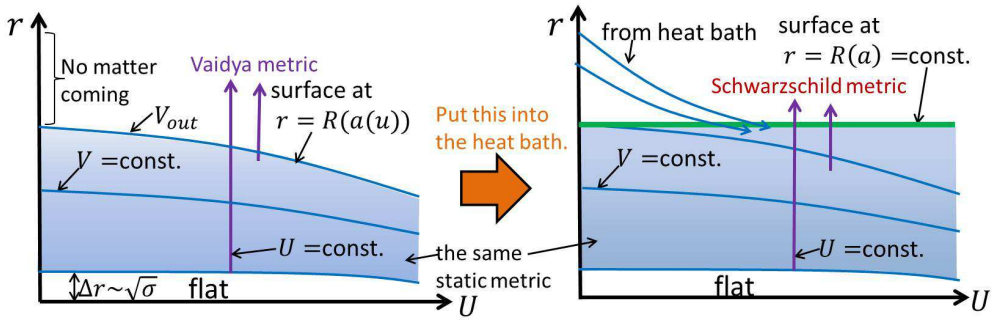}
\caption{The boundary conditions. 
Left: The evaporating black hole in the vacuum. 
Right: The stationary black hole in the heat bath.}
\label{boundary}
\end{center}
\end{figure}

We start with the boundary conditions. 
See Fig.\ref{boundary}. 
We first note that the region around $r=0$ is kept to be a flat space.  
This is because the initial collapsing matter came from infinity with a dilute distribution. 
Then, the region inside the innermost shell in Fig.\ref{multi} is flat due to the spherical symmetry, 
and it is almost frozen in time by the large redshift as in \Ref{eva_metric}
\footnote{We will check the validity of \Ref{Tr} later.
Indeed, \Ref{Tr} becomes almost flat at $r\sim \sqrt{\s}$, 
and can be connected to the flat spacetime.}. 
Thus, the boundary conditions for $\T$ are given by 
\begin{equation}\lb{bound_EMT_in}
\bra T_{\mu\nu}\ket|_{r\sim 0}=0.
\end{equation}
Note that this should be applied to both the 
evaporating and stationary black holes, 
because at any rate black holes have been formed by collapse of matters.

\subsubsection{Employing $\N^\mu\bra T_{\mu\nu}\ket =0$}
Now, we combine the energy-momentum conservation with the assumption \Ref{T_UV}.
Under \Ref{T_UV}, \Ref{formula} becomes 
\begin{equation}\lb{con_U2}
\p_r (r^2\bra T_{UU}\ket) = \f{1}{2}r e^{\vp} \bra T^\mu{}_\mu \ket.
\end{equation}
Integrating this from $0$ to $r$ for $\sqrt{\s}\ll r \leq R(a)$, 
we have
\begin{align}\lb{integ}
r^2\bra T_{UU}\ket-(r^2\bra T_{UU}\ket)|_{r=0}& = \f{1}{2}\int^r_0dr'r'e^{\vp(r')}\bra T^\mu{}_\mu (r')\ket 
= \s \int^r_0dr'\f{e^{\f{r'^2}{2\s}}}{r'}\bra T^\mu{}_\mu (r')\ket\nn \\ 
 &= \s e^{\f{r^2}{2\s}}\int^r_0dr'\f{e^{-\f{r^2-r'^2}{2\s}}}{r'}\bra T^\mu{}_\mu (r')\ket \nn\\
 &\approx \f{\s}{r}e^{\f{r^2}{2\s}}\bra T^\mu{}_\mu (r)\ket \int^r_0dr'e^{-\f{r}{\s}(r-r')} \nonumber \\
 &\approx \f{\s^2}{r^2}e^{\f{r^2}{2\s}}\bra T^\mu{}_\mu (r)\ket.  
\end{align}
Here, at the first line, we have used \Ref{UV2} and \Ref{AB_sta}; 
at the third line, we have assumed that $\bra T^\mu{}_\mu (r)\ket$ does not change as rapidly as $e^{\f{r^2}{2\s}}$, 
which will be checked soon, and used $e^{-\f{1}{2\s}(r+r')(r-r')}\approx e^{-\f{r}{\s}(r-r')}$, 
since the largest contribution comes from $r'\sim r$; 
at the final line, we have omitted the term proportional to $e^{-\f{r^2}{\s}}$ for $r\gg \sqrt{\s}$. 
Finally, using the boundary condition \Ref{bound_EMT_in}, 
we have 
\begin{equation}\lb{TT_theta}
\bra T_{UU} \ket=\bra T_{VV} \ket=\f{\s^2}{r^4}e^{\f{r^2}{2\s}}\bra T^\mu{}_\mu\ket.
\end{equation}

On the other hand, under the assumption \Ref{T_UV}, \Ref{trace2} leads to 
\begin{equation}\lb{ano_theta}
\bra T^\th{}_\th \ket = \f{1}{2} \bra T^\mu{}_\mu \ket.
\end{equation}
Thus, all the components of $\T$ are determined by $\Tt$.

\subsubsection{$\Tt$ from the 4D Weyl anomaly}
In the case of conformal matters, 
$\Tt$ is provided by the 4D Weyl anomaly once the metric is given \ci{Christensen,Duff,BD,Deser}: 
\begin{equation}\lb{anomaly}
\bra T^\mu{}_\mu \ket=\hbar c_{\rm w} {\cal F}- \hbar a_{\rm w} {\cal G}, 
\end{equation}
where ${\cal F}\equiv C_{\mu\nu\a\b}C^{\mu\nu\a\b}$ and ${\cal G}\equiv R_{\mu\nu\a\b}R^{\mu\nu\a\b}-4R_{\mu\nu}R^{\mu\nu}+R^2$
\footnote{We assume that the coefficients of the higher-curvature terms 
in the effective action are renormalized to order 1. 
However, $c_{\rm w}$ and $a_{\rm w}$ are proportional to the degrees of freedom $N$ 
because they are not canceled by counterterms \ci{BD}.
Therefore, we can ignore the contributions from the higher curvature terms 
if $N \gg 1$. }.
For the metric \Ref{Tr}, ${\cal F}$ and ${\cal G}$ are calculated as
\begin{align}\lb{F_G}
{\cal F} &= \f{A'^4}{12 B^2}+\cdots =\f{1}{3\s^2}+O\l(\f{1}{\s r^2}\r)\nn\\
{\cal G}  &= -\f{2A'^2}{r^2B}+\cdots=O\l(\f{1}{\s r^2}\r). 
\end{align}
Therefore, only the $c$-coefficient remains for $r\gg \sqrt{\s}$, 
and we obtain 
\begin{equation}\lb{ano_value}
\Tt=\f{\hbar \cw}{3\s^2}, 
\end{equation}
which is constant and consistent with the assumption made in \Ref{integ}.

Thus, \Ref{TT_theta} and \Ref{ano_theta} are fixed as, respectively,
\begin{equation}\lb{T_UU_VV}
\bra T_{UU} \ket=\bra T_{VV} \ket=\f{\hbar c_{\rm w}}{3 r^4} e^{\f{r^2}{2\s}},
\end{equation}
and 
\begin{equation}\lb{T_th}
\bra T^\th{}_\th \ket = \f{\hbar c_{\rm w}}{6\s^2},
\end{equation}
which means that the 4D Weyl anomaly provides the angular pressure \ci{KY1,KY2}
\footnote{See e.g. \ci{Kim} for another application of the 4D Weyl anomaly to black holes.}. 

\subsection{The self-consistent equation}
Now we can obtain the condition that the self-consistent equation \Ref{Einstein} holds, as follows.
From \Ref{T_UV}, \Ref{T_UU_VV} and \Ref{T_th}, we have
\begin{equation}\lb{T_inside}
-\bra T^T{}_T \ket =\bra T^r{}_r \ket = \f{\hbar c_{\rm w}}{3\s}\f{1}{r^2},~~\bra T^\th{}_\th \ket = \f{\hbar c_{\rm w}}{6\s^2},
\end{equation}
where we have used \Ref{UV_T}. 
On the other hand, the Einstein tensor for the metric \Ref{Tr} is calculated as 
\begin{equation}\lb{G_Tr}
-G^T{}_T=G^r{}_r=\f{1}{r^2},~~~G^\th{}_\th=\f{1}{2\s}. 
\end{equation}
Comparing \Ref{T_inside} and \Ref{G_Tr}, we conclude that 
\Ref{Einstein} is satisfied if we identify 
\begin{equation}\lb{s_value}
\s = \f{8\pi l_p^2 c_{\rm w}}{3}. 
\end{equation}
We note that 
the dominant energy condition \ci{Poisson} is violated, $-\bra T^T{}_T \ket \ll \bra T^\th{}_\th \ket$, 
and that the interior is not a fluid in the sense $\bra T^r{}_r \ket \ll \bra T^\th{}_\th \ket$ \ci{KMY,KY1,KY2}. 

We can check the validity of the classical gravity in \Ref{Einstein}. 
Indeed, in the macroscopic region $(r>l_p)$, all the invariants for \Ref{Tr} are of order $\sim \f{1}{\s}$: 
\begin{equation}\lb{curve}
R,~\sqrt{R_{\mu\nu}R^{\mu\nu}},~\sqrt{R_{\mu\nu\a\b}R^{\mu\nu\a\b}}\sim \f{1}{ \s}\sim \f{1}{l_p^2 c_{\rm w}}.
\end{equation}
They are smaller than the Planck scale if 
\begin{equation}\lb{large_c}
c_{\rm w}\gg 1
\end{equation}
is satisfied. 
Therefore, macroscopic black holes $(a \gg l_p)$ can be described by the ordinary field theory. 
We do not need to consider quantum gravity except for the very small region $(r\sim l_p)$ 
or the last moment of the evaporation. 
\Ref{Tr} can be trusted  for $r\gtrsim \sqrt{\s}$. 

\subsection{Evaluation of $\bra T_{\mu\nu} \ket$ outside the black hole}\lb{s:T_outside}
In this subsection we investigate $\T$ in the outside region, $r>R(a)$, 
for both the evaporating and the stationary black holes. 
\subsubsection{The evaporating black hole}
First we consider the evaporating back hole \Ref{eva_metric}.
Although we don't assume the static condition \Ref{T_static}, 
we use a similar argument to the previous subsection. 
We first identify the boundary conditions. 
In the left of Fig.\ref{boundary}, 
no ingoing matter comes after the collapsing matter at $U=-\infty$. 
Therefore, the boundary condition for the ingoing energy $\bra T_{VV} \ket$ is given by 
\begin{equation}\lb{bc_V}
\bra T_{VV} \ket|_{U=-\infty}=0~~{\rm for}~~V> V_{out},
\end{equation}
where $V_{out}$ labels the outermost shell.
On the other hand, as we have shown in \Ref{T_UU_VV}, 
the outgoing energy at the surface $r=R(a(U))$ is given by
\begin{equation}\lb{bc_U}
\bra T_{UU} \ket|_{V=V_{out}}= \f{\hbar c_{\rm w}}{3R(a(U))^4} ~~{\rm for}~~U\geq U_{0}.
\end{equation}
Here we have identified $U$ in \Ref{UV} with $u$ in \Ref{Vaidya} 
so that $A=\f{r^2-R(a)^2}{2\s}$ as in \Ref{eva_metric}. 
$U_{0}$ characterizes the time at which the outermost shell gets sufficiently close to $R(a(U))$ 
and starts to emit the radiation. 

Using these boundary conditions and the conservation laws \Ref{con_Ut} and \Ref{con_Vt} with the assumption \Ref{T_UV}, 
we obtain (see Appendix \ref{ap:EMT_Vaidya} for the derivation.)
\begin{align}
\lb{Vaidya_U2}
r^2 \bra T_{UU}\ket&= \f{\hbar c_{\rm w}}{3R(a(U))^2}+\f{1}{2}\int^{r(U,V)}_{R(a(U)),U={\rm const.}} dr (r-a(U))  \bra T^\mu{}_\mu\ket,\\
\lb{Vaidya_V2}
r^2 \bra T_{VV}\ket&= - \int^U_{-\infty} dU' r (\p_Vr)^2 \bra T^\mu{}_\mu \ket.
\end{align}
Next, we evaluate $\bra T^\mu{}_\mu \ket$ from \Ref{anomaly}.  
For the metric \Ref{eva_metric} for $r>R(a(u))$, 
we have ${\cal F}={\cal G}=\f{12 a(U)^2}{r^6}$ and obtain 
\begin{equation}\lb{Vaidya_p}
\Tt=12\hbar (c_{\rm w}-a_{\rm w})\f{a(U)^2}{r^6},
\end{equation}
which gives $\Tth$ through \Ref{ano_theta}.
From \Ref{Vaidya_U2} and \Ref{Vaidya_p}, 
we obtain
\begin{equation}\lb{Vaidya_UU}
r^2 \bra T_{UU}\ket\approx\hbar\l(\f{c_{\rm w}}{3} +\f{3(c_{\rm w}-a_{\rm w})}{10} \r)\f{1}{a(u)^2}
+6\hbar (c_{\rm w}-a_{\rm w}) a(u)^2 \l(-\f{1}{4r^4} + \f{a(u)}{5r^5}\r),
\end{equation}
where $R(a)\approx a$ has been used. 
On the other hand, \Ref{Vaidya_V2} cannot be evaluated explicitly due to the time dependence of $a(U)$. 
Here, in order to estimate its order,  we assume that $a(U)$ is approximately constant.  
Then, we can have (see Appendix \ref{ap:EMT_Vaidya})
\begin{equation}\lb{Vaidya_VV}
r^2 \bra T_{VV}\ket \sim \hbar (c_{\rm w}-a_{\rm w}) a^2 \l(-\f{1}{4r^4} + \f{a}{5r^5}\r).
\end{equation}

Note here that the anomaly leads to particle creation even outside the black hole. 
The sign of $c_{\rm w}-a_{\rm w}$ depends on the kind of field \ci{BD}. 
For example, it is positive for a massless scalar field, and it is negative for a massless vector field
\footnote{However, $\f{c_{\rm w}}{3} +\f{3(c_{\rm w}-a_{\rm w})}{10} >0$ holds 
for any kind of massless fields \ci{BD}, and $\bra T_{UU} \ket$ is always positive at infinity.
Here the boundary condition \Ref{bc_U} plays an important role. 
Later we will discuss the origin of the radiation more closely.
}. 
When $c_{\rm w}-a_{\rm w}>0$, 
\Ref{Vaidya_UU} indicates that 
the outgoing radiation increases by the amount $\f{3\hbar(c_{\rm w}-a_{\rm w})}{10} \f{1}{a(U)^2}$ 
as it goes to infinity from the surface. 
On the other hand, from \Ref{Vaidya_VV},  
we can see that 
the negative ingoing energy is created \ci{Christensen,BD,DFU}. 

Now we check the self-consistent equation \Ref{Einstein}. 
First, from \Ref{Vaidya_p}, \Ref{Vaidya_UU} and \Ref{Vaidya_VV}, 
we can see that $\bra T_{\mu\nu} \ket \sim \f{1}{a^4}$ at $r\sim a$, 
which represents the energy-momentum of the radiation around the black hole 
as in the Stefan-Boltzmann law $\sim T_H^4$. 
The amount of energy in the region around the black hole with the volume $V\sim a^3$ 
is estimated as $\T V \sim \f{1}{a}$, which is much smaller than
the mass of the black hole itself, $M=\f{a}{2G}$. 
In this sense, $\T$ is negligible: 
\begin{equation}\lb{T_outside}
\bra T_{\mu\nu}\ket \approx 0,
\end{equation}
and the region outside the black hole is described by vacuum-like solutions 
such as the Vaidya metric or the Schwarzschild metric. 

\begin{figure}[h]
 \begin{center}
 \includegraphics*[scale=0.18]{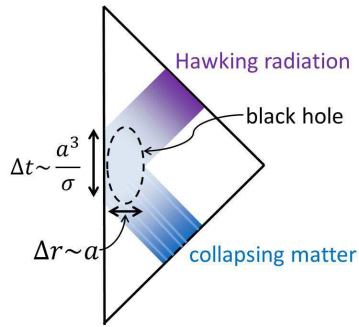}
 \caption{The Penrose diagram of the evaporating black hole described by the self-consistent solution \Ref{eva_metric}.}
 \label{Penrose}
 \end{center}
 \end{figure}

We have seen so far that 
the metric \Ref{eva_metric} is the self-consistent solution describing the whole spacetime 
of the evaporating black hole. 
There is no horizon or singularity, 
but this object is the black hole in quantum mechanics (see Fig.\ref{Penrose}). 

\subsubsection{The stationary black hole}
Next we consider the stationary black hole in the heat bath \Ref{sta_metric}. 
This time we assume \Ref{T_static} in addition to \Ref{T_UV}, 
and use \Ref{con_U2}. 
We start with examining the boundary condition. See the right of Fig.\ref{boundary}. 
Because the system is stationary, the surface is fixed at $r=R(a)=$const, 
and there the ingoing and outgoing energy flows are balanced as
\begin{equation}\lb{bc_stat}
\bra T_{UU} \ket|_{r=R(a)}=\bra T_{VV} \ket|_{r=R(a)}=\f{\hbar c_{\rm w}}{3 R(a)^4}. 
\end{equation}
Here we have used \Ref{T_UU_VV} and chosen the overall time scale as in \Ref{sta_metric}, $A(r)=\f{r^2-R(a)^2}{2\s}$. 

Then, we calculate $\Tt$ from \Ref{ano_theta} 
and obtain the same value as \Ref{Vaidya_p} except for $a=$const.
We can evaluate $\bra T_{UU} \ket$ from \Ref{con_U2} with \Ref{bc_stat}, 
and find that $\bra T_{UU} \ket=\bra T_{VV} \ket$ is given by \Ref{Vaidya_UU} with $a=$const.

Now we study the self-consistent equation. 
Because we have the same order of $\T$ as in the case of the evaporating black hole, 
we can follow the same reasoning for \Ref{T_outside}. 
That is, $\T$ is negligible, and the metric outside the black hole is close to 
the Schwarzschild metric. 

\section{Generalization}\lb{s:Gen}
We have assumed so far that the radiation emitted from each 
shell flows to infinity without reflection, 
which is expressed by \Ref{T_UV}. 
For a more realistic description, 
however, this assumption should be removed.

First we discuss what $\bra T_{UV} \ket \neq0$ means. 
In the $(U,V)$ coordinates \Ref{UV}, this is equivalent to 
the nonzero trace in the 2-dimensional part $(U,V)$: 
\begin{equation}\lb{2d_tr}
\bra T^a{}_a \ket\equiv \bra T^U{}_U \ket+\bra T^V{}_V \ket=2g^{UV}\bra T_{UV} \ket. 
\end{equation}
In a $(t,r)$ coordinate system, in which the metric is diagonal, this is expressed as
\begin{equation}\lb{2d_tr2}
\bra T^a{}_a \ket= \bra T^t{}_t \ket+\bra T^r{}_r \ket.
\end{equation}
In other words, $\bra T_{UV} \ket=0$ is equivalent to $-\bra T^t{}_t \ket=\bra T^r{}_r\ket$, 
which is indeed satisfied by the previous self-consistent solution as in \Ref{T_inside}. 
Therefore, we characterize $\bra T_{UV} \ket \neq0$ by introducing a function $f(t,r)$ such that
\begin{equation}\lb{f}
\f{\bra T^r{}_r \ket}{-\bra T^t{}_t \ket} \equiv \f{1-f}{1+f}.
\end{equation}
$f=0$ corresponds to $-\bra T^t{}_t \ket=\bra T^r{}_r\ket$.
Here if we require $\bra T^r{}_r\ket \geq 0$ and $-\bra T^t{}_t \ket> 0$, 
$f$ must satisfy $|f|\leq 1$. 
In the following arguments, we assume that the matters are conformal.

\subsection{Determination of the interior metric}
For simplicity, we consider a stationary black hole in the heat bath. 
More precisely, we describe the exterior by the Schwarzschild metric \Ref{Sch}, 
and parametrize the interior metric by \Ref{Tr_ansazt} \ci{KY1}. 
Then, we assume that $\bra T_{\mu\nu} \ket$ is static and satisfies \Ref{T_static}. 
Our program is to fix two functions $A(r)$ and $B(r)$ by two equations. 

The first equation comes from \Ref{f}. 
Once $f(r)$ is given, 
we rewrite the relation \Ref{f}, 
by using the self-consistent equation \Ref{Einstein} for the ansazt \Ref{Tr_ansazt}, as 
\begin{equation}\lb{f3}
\f{2}{1+f} =\f{G^r{}_r}{-G^t{}_t}+1=\f{r\p_r A}{B-1+r\p_r \log B}.
\end{equation}
In order to build the second equation, 
we apply the Weyl anomaly formula \Ref{anomaly} 
to the trace of \Ref{Einstein}: 
\begin{equation}\lb{anomaly_eq}
G^\mu{}_\mu=8\pi G \bra T^\mu{}_\mu \ket=\g {\cal F}- \a {\cal G},
\end{equation}
where we have introduced the notations $\g\equiv 8\pi G \hbar c_{\rm w}$ and $\a\equiv 8\pi G \hbar a_{\rm w}$. 

Here, we assume that for $r\gg l_p$, $A(r)$ and $B(r)$ are large quantities of the same order 
as expected from \Ref{AB_sta}:
\begin{equation}\lb{AB}
A(r)\sim B(r)\gg 1~.
\end{equation}
Then, the first equation \Ref{f3} becomes approximately 
\begin{equation}\lb{self1}
A'=\f{2B}{(1+f)r},  
\end{equation}
where $A'=\p_r A$ and we have used $B\gg 1,r\p_r \log B$. 
Next, in order to examine what terms dominate in \Ref{anomaly_eq} for $r\gg l_p$, 
we replace $A$, $B$, and $r$ with $\mu A$, $\mu B$, and $\sqrt{\mu} r$, respectively, 
and pick up the terms with the highest powers of $\mu$. 
Then, we have 
\begin{equation}\lb{anomaly_eq2}
\f{A'^2}{2B}+\cdots=\g\l(\f{A'^4}{12 B^2}+\cdots  \r)-\a \l(-\mu^{-1}\f{2A'^2}{r^2B}+\cdots \r).
\end{equation}
Therefore, in the leading order of $r$, \Ref{anomaly_eq} becomes $\f{A'^2}{2B}=\g\f{A'^4}{12 B^2}$, that is, 
\begin{equation}\lb{self2}
B=\f{\g}{6}A'^2 ~.
\end{equation}

It is natural to expect that the dimensionless function $f(r)$  
is a constant for conformal fields \ci{KY1}: 
\begin{equation}\lb{f_cft}
f(r)={\rm const.}
\end{equation}
Then, from \Ref{self1}, \Ref{self2} and \Ref{f_cft}, we obtain 
\begin{equation}\lb{AB_f}
A=\f{r^2}{2(1+f)\s_f},~~~B=\f{r^2}{2\s_f},
\end{equation}
where we have defined 
\begin{equation}\lb{sigma_f}
\s_f \equiv \f{8\pi l_p^2 c_{\rm w}}{3(1+f)^2}. 
\end{equation}
Thus, the interior metric is determined as 
\begin{equation}\lb{Tr_f}
d s ^2 = -\f{2\s_f}{r^2} e^{\f{r^2}{2(1+f)\s_f}} d T^2 
+ \f{r^2}{2\s_f} d r^2 + r^2 d \Omega^2. 
\end{equation}
Indeed, this is a generalization of \Ref{Tr} 
because \Ref{sigma_f} and \Ref{Tr_f} become \Ref{s_value} and \Ref{Tr}, respectively, if we set $f=0$. 
Redefining the overall scale of time and connecting the metric with the Schwarzschild metric,
we reach the generalized metric for the stationary black hole: 
\begin{equation}\lb{gen_metric}
d s^2=\begin{cases}
-\f{2\s_f}{r^2} e^{- \f{R(a)^2-r^2}{2(1+f)\s_f}} d t^2 + \f{r^2}{2\s_f} d r^2 + r^2 d \Omega^2,~~{\rm for}~~r\leq R(a)~,\\
 - \f{r-a}{r}d t^2 + \f{r}{r-a}d r^2 + r^2 d \Omega^2,~~{\rm for}~~r\geq R(a)~,
\end{cases}
\end{equation}
where $R(a)=a+\f{2\s_f}{a}$. 
The metric for the evaporating one is obtained 
with the outside metric replaced by the Vaidya metric \Ref{Vaidya}. 

\subsection{Check of the self-consistent equation}
As in section \ref{s:EMT},
we now evaluate $\bra T_{\mu\nu} \ket$ in the metric \Ref{gen_metric}, 
and check the self-consistent equation.
Because we assume that $\bra T_{\mu\nu} \ket$ is static, 
we have to determine three functions of $r$: 
$\bra T_{UU} \ket = \bra T_{VV} \ket$, $\bra T_{UV} \ket$, and $\bra T^\th{}_\th \ket$. 

\subsubsection{Evaluation of $\bra T_{\mu\nu} \ket$ inside the black hole}
First we determine $\bra T_{\mu\nu} \ket$ in the interior metric \Ref{Tr_f}, 
which can be expressed by \Ref{UV} with \Ref{AB_f}. 
We assume \Ref{f_cft} and express the relation \Ref{f} as 
\begin{equation}\lb{f_rel}
\bra T_{UV} \ket =f \bra T_{UU} \ket,
\end{equation}
where we have used \Ref{UV_T}.
Thus, only $\bra T_{UU} \ket$ and $\bra T^\th{}_\th \ket$ are left as unknown functions. 

We then substitute \Ref{f_rel} to \Ref{formula} and obtain
\begin{equation}\lb{eq_f1}
\p_r ((f-1)r^2\bra T_{UU}\ket) - f\l(\p_r \vp - \f{2}{r} \r) (r^2\bra T_{UU}\ket)
= -\f{1}{2}r e^{\vp} \bra T^\mu{}_\mu \ket.
\end{equation}
Using \Ref{f_cft}, \Ref{AB_f} and $\p _r \vp \approx \p _r A =\f{r}{(1+f)\s_f} \gg \f{2}{r}$ for $r\gg l_p$, 
we reach 
\begin{equation}\lb{eq_f1}
\p_r (r^2\bra T_{UU}\ket) + \f{f}{(1-f^2) \s_f}r (r^2\bra T_{UU}\ket)
= \f{\s_f}{(1-f)r}e^{\f{r^2}{2(1+f)\s_f}} \bra T^\mu{}_\mu \ket.
\end{equation}
The solution can be expressed as 
\begin{equation}
r^2\bra T_{UU}(r)\ket=C(r)e^{-\f{f}{2(1-f^2)\s_f}r^2},
\end{equation}
where $C(r)$ satisfies
\begin{equation}
\p_r C =\f{\s_f}{(1-f)r} e^{\f{r^2}{2(1-f^2)\s_f}} \bra T^\mu{}_\mu \ket.
\end{equation}
This equation can be solved easily as 
\begin{align}
C(r)-C(0) &= \f{\s_f}{(1-f)} \int^r_0 dr' \f{1}{r'}e^{\f{r'^2}{2(1-f^2)\s_f}} \bra T^\mu{}_\mu(r') \ket \nn\\
 &\approx \f{(1+f)\s_f^2}{r^2} e^{\f{r^2}{2(1-f^2)\s_f}} \bra T^\mu{}_\mu(r) \ket,
\end{align}
where we have employed almost the same technique as in \Ref{integ}. 
Here the boundary condition \Ref{bound_EMT_in} means $C(0)=0$. 
Then, we reach
\begin{equation}\lb{eq_f2_pre}
r^2\bra T_{UU}(r)\ket=\f{(1+f)\s_f^2}{r^2} e^{\f{r^2}{2(1+f)\s_f}} \bra T^\mu{}_\mu(r) \ket.
\end{equation}

Applying the Weyl anomaly formula \Ref{anomaly} to the metric \Ref{Tr_f} 
and using the same estimation as \Ref{F_G}, we have 
\begin{equation}\lb{anomaly_f}
\bra T^\mu{}_\mu \ket= \f{\hbar c_{\rm w}}{3(1+f)^4 \s_f^2} = \f{3}{(8\pi)^2 G l_p^2 c_{\rm w}},
\end{equation}
where at the second equality we have used \Ref{sigma_f}
\footnote{We note that $\Tt$ is independent of $f$.}. 
Substituting this into \Ref{eq_f2_pre}, we obtain 
\begin{equation}\lb{eq_f2}
r^2\bra T_{UU}(r)\ket=\f{\hbar c_{\rm w}}{3 (1+f)^3 r^2} e^{\f{r^2}{2(1+f)\s_f}},
\end{equation}
which reduces to \Ref{T_UU_VV} if $f=0$. 
Then, from \Ref{trace2}, \Ref{f_rel} and \Ref{eq_f2}, we obtain 
\begin{equation}\lb{pth_f}
\bra T^\th{}_\th \ket = \f{3}{2(8\pi)^2 G l_p^2 c_{\rm w}}
+\f{f}{8\pi G (1+f)r^2} \approx \f{3}{2(8\pi)^2 G l_p^2 c_{\rm w}}.
\end{equation}

Now we can check the self-consistent equation \Ref{Einstein} explicitly. 
Using \Ref{f_rel}, \Ref{eq_f2} and \Ref{UV_T}, 
we have 
\begin{align}\lb{rho_f}
-\bra T^t{}_t \ket &= \f{\hbar c_{\rm w}}{3\s_f (1+f)^2 r^2 }=\f{1}{8\pi G r^2}  \\
\lb{pr_f}
\bra T^r{}_r \ket &= \f{\hbar c_{\rm w}(1-f)}{3\s_f (1+f)^3 r^2 }=\f{1}{8\pi G r^2}\f{1-f}{1+f} 
\end{align}
where at the second equality we have used \Ref{sigma_f}.
On the other hand, we have for the metric \Ref{Tr_f} 
\begin{equation}\lb{G_f}
-G^t{}_t= \f{1}{r^2},~~~G^r{}_r = \f{1}{r^2}\f{1-f}{1+f},~~~G^\th{}_\th=\f{1}{2(1+f)^2 \s_f} = \f{3}{16\pi l_p^2 \cw}.
\end{equation}
Comparing \Ref{pth_f}, \Ref{rho_f} and \Ref{pr_f} with \Ref{G_f}, 
we find that \Ref{Einstein} is indeed satisfied. 

Finally, we see that the quantum fluctuation of gravity is small also in the general case. 
In fact, the invariants of \Ref{Tr_f} are given by 
\begin{equation}\lb{curve2}
R,~\sqrt{R_{\mu\nu}R^{\mu\nu}},~\sqrt{R_{\mu\nu\a\b}R^{\mu\nu\a\b}}\sim \f{1}{(1+f)^2 \s_f} \sim \f{1}{l_p^2 c_{\rm w}},
\end{equation}
where \Ref{sigma_f} has been used. 
They are small compared with the Planck scale, 
and therefore the fluctuation is small if \Ref{large_c} is satisfied. 

\subsubsection{Evaluation of $\bra T_{\mu\nu} \ket$ outside the black hole}
Next we consider the outside region, $r>R(a)$, of the metric \Ref{gen_metric}. 
As we have seen in the previous section, $\bra T_{\mu\nu} \ket$
outside the black hole is so small that the modification
from the Schwarzschild or Vaidya metric is negligible,
although the precise condition to fix $\bra T_{\mu\nu} \ket$ is not
known.
In this subsection, as a simple example, we fix $\TU$ by hand and determine $\TUV$. 
Then, we show that the region outside the black hole can be described approximately by the Schwarzschild metric. 

We assume  
\begin{equation}\lb{given_UU}
\bra T_{UU} (r) \ket= \f{\hbar c_{\rm w}}{3 (1+f)^3 R(a)^2}\f{1}{r^2},
\end{equation}
where $f$ is a constant given by \Ref{f_cft}. 
This means that the total flux emitted from the surface at $r=R(a)$ is kept outside (see \Ref{eq_f2} for $A=\f{r^2 - R(a)^2}{2(1+f)\s_f}$)
while the other effects (such as particle creation outside the black hole by the anomaly in subsection \ref{s:T_outside}) 
do not contribute to $\TU$. 
Furthermore, 
we take for simplicity
\begin{equation}\lb{surface_UV}
\bra T_{UV}\ket|_{r=R(a)}= 0, 
\end{equation}
as the boundary condition. 
We note that \Ref{given_UU} and \Ref{surface_UV} are not given by some principle but chosen by hand as an example. 

Then, the first term in the right hand side of \Ref{formula} vanishes 
while the second term is given through the Weyl anomaly by \Ref{Vaidya_p} with $a=$const. 
Solving \Ref{formula} with the method of variation of constants under \Ref{surface_UV}, 
we obtain
\footnote{
For given $\bra T_{UU} \ket$ and $\bra T^\mu{}_\mu \ket$, 
we solve \Ref{formula} with respect to $r^2 \bra T_{UV} \ket$ and have 
$r^2 \bra T_{UV} (r)\ket=D(r) e^{\varphi-2 \log r}= D(r) \f{r-a}{r^3}$,
where $e^\vp = \f{r-a}{r}$ has been used. 
Then, $D(r)$ satisfies 
$\p_rD= \f{r^3}{r-a} \p_r (r^2 \bra T_{UU} (r)\ket)- \f{1}{2} r^3 \bra T^\mu{}_\mu \ket$.
Applying \Ref{given_UU} and \Ref{Vaidya_p} to this 
and integrating it from $R(a)$ to $r$, we obtain \Ref{Sc_UV2} if \Ref{surface_UV} is considered.}
\begin{equation}\lb{Sc_UV2}
\bra T_{UV} (r)\ket=3\hbar (c_{\rm w}-a_{\rm w}) a^2 \l( \f{1}{r^2}-\f{1}{R(a)^2} \r)\f{r-a}{r^5}. 
\end{equation}
This behaves $\sim \f{a^2}{r^4}$ for $r\gg a$, 
which decreases faster than \Ref{given_UU}, and does not contribute to the flux at infinity. 
Using \Ref{Vaidya_p} and \Ref{Sc_UV2}, we can evaluate $\bra T^\th{}_\th \ket$ through \Ref{trace2} as
\begin{equation}
\bra T^\th{}_\th \ket = 6\hbar (c_{\rm w}-a_{\rm w}) \f{a^2}{r^4} \l( \f{2}{r^2}-\f{1}{R(a)^2} \r). 
\end{equation}

Thus, $\T\sim \f{1}{a^4}$ around $r\sim a$, 
and we can regard $\T\approx 0$ by the same reasoning for \Ref{T_outside}. 
Therefore, \Ref{Einstein} is satisfied by \Ref{gen_metric}.

\section{Hawking radiation}\lb{s:radiation}
In this section we discuss how close the object that we are considering
is to the black hole in the conventional picture.
\subsection{Amount of the radiation}
First we show that the object emits the same amount of radiation as 
the conventional black hole. 
We prove that the energy flux at $r$ is given by 
\begin{equation}\lb{J_rad}
J(r)=\f{4\pi \hbar c_{\rm w}}{3(1+f)^2r^2}=\f{\s_f}{2G r^2},
\end{equation}
where $J$ is the energy passing through the ingoing spherical null surface at $r$ 
per unit time.  
Here the time is ``the local time at $r$'' such as $u_i$ in \Ref{Vaidya_i} 
for the multi-shell model. 
(Then, \Ref{J_rad} agrees with the right hand side of \Ref{da_i}.) 
More precisely, we define $J$ by \footnote{
We can see that this definition is consistent with the concept of $J$, as follows. 
To do that, we first note that \Ref{da_i} suggests $u_i$ as the natural time 
for description of the evaporation of each shell, 
and that in the continuum limit the redshift factor between $U$ and $u_i$ is $e^{\f{A}{2}}$, 
as \Ref{evalu2} shows. 
Then, we introduce the energy-momentum vector observed by $\bm u $ as 
$P^\mu \equiv - \bra T^\mu{}_\nu \ket u^\nu$. 
Here $\bm u $ is the 4-vector with time $u_i$, 
which is defined by 
$\bm u \equiv e^{-\f{A}{2}}\l(\f{\p}{\p U} \r)_r= 
e^{-\f{A}{2}}\l[ \l(\f{\p}{\p U} \r)_V+ \l(\f{\p}{\p V} \r)_U\r]$. 
Here we have used \Ref{T} and \Ref{UV_T}. 
Thus, we can identify $J$ with 
$J= 4\pi r^2 (- P^\mu k_\mu)$, 
where $\bm k \equiv e^{-\f{A}{2}} \l(\f{\p}{\p U} \r)_V$
is the ingoing null vector along the shell. }
\begin{equation}\lb{J}
J(r)\equiv 
4\pi r^2 e^{-A} (\bra T_{UU}\ket + \bra T_{UV}\ket ). 
\end{equation}
We can easily show that \Ref{J} becomes \Ref{J_rad}
by using \Ref{AB_f}, \Ref{f_rel} and \Ref{eq_f2}.
Note that \Ref{J_rad} means that 
the $c$-coefficient determines the intensity of the Hawking radiation 
and the effect of $f$ is to decrease the flux \ci{KY1,KY2}. 

Now we apply \Ref{J_rad} to the surface $r=R(a)$, 
and obtain the energy flux emitted by the object: 
\begin{equation}\lb{J_tot}
J(R(a))=\f{\s_f}{2G R(a)^2},
\end{equation}
which agrees with the amount of the radiation emitted 
by the black hole in the conventional picture.

Here we point out that
we can obtain the energy spectrum of the radiation
by solving the wave equation in the metric \Ref{eva_metric} under the eikonal approximation. Indeed
it turns out to be the Planck-like distribution with the Hawking temperature \ci{KMY, KY2}.

\subsection{Insensitivity to the detail of the initial wave function}
Next we argue that 
the expectation value of the energy momentum tensor
is determined by the overall geometry, and does not depend on the
detail of the initial wave function.
To see this, 
we start with reexamining the analysis \Ref{integ} of $\N^\mu \bra T_{\mu U} \ket=0$. 
If we integrate it from $r=r_0$ instead of $r=0$, we have 
\begin{equation}\lb{integ_re}
r^2 \bra T_{UU} \ket = (r^2 \bra T_{UU} \ket)|_{r_0} + 
\f{\s^2}{r^2} \bra T^\mu{}_\mu (r)\ket e^{\f{r^2}{2\s}} (1-e^{-\f{r}{\s}(r-r_0)}).
\end{equation}
Here, the last term vanishes for such $r_0$ that $\f{r}{\s}(r-r_0)\gg 1$,
and the first term is negligible 
unless it is as large as $O(r^{-2}e^{\f{r^2}{2\s}})$.
Thus, even if we do not use the boundary condition \Ref{bound_EMT_in}, 
we obtain the same result \Ref{TT_theta}. 

This indicates that the amount of the radiation is determined universally by the geometry. 
Indeed as is shown in \Ref{TT_theta}, 
$\TU$ is produced at each point in the interior through the 4D Weyl anomaly \Ref{anomaly}, 
which is independent of the state but is determined by the metric \Ref{Tr}. 
Furthermore, while we have assumed the configuration \Ref{R_i} to obtain the metric \Ref{Tr}, 
it has been shown by \ci{Ho3} that 
\Ref{Tr} is asymptotically reached from any initial distribution of mass and velocity of the matter. 
In this sense
the radiation occurs universally in collapsing processes,
whose amount is given by \Ref{J_tot}. 

Here we emphasize that the 4D Weyl anomaly plays a crucial role 
in our picture of black holes. 
As \Ref{ano_theta} shows, 
the anomaly induces the strong angular pressure \Ref{T_th} \ci{DFU,CGHS,RST,Wilczek,Iso}. 
It is so strong in the metric \Ref{Tr} that 
the object can be stable against the strong gravitational force
\footnote{We can see explicitly this by constructing the Tolman-Oppenheimer-Volkoff equation 
with $\bra T^r{}_r \ket \neq \Tth$ and using $-\bra T^t{}_t \ket, \bra T^r{}_r \ket \ll \Tth$.}
\footnote{See also \ci{Abedi}. }.

\subsection{Fate of the incoming matter}
Finally we discuss the information problem.
In our picture the matter fields simply propagate in the background metric as in the ordinary
quantum field theory on curved spacetime, and nothing special
happens during the time evolution.
Therefore, it is natural to expect that 
the collapsing matter itself eventually comes back as the radiation. 

Indeed, we can get a clue to this by a simple analysis \ci{KY2}. 
Suppose that a particle with energy $\sim \f{\hbar}{a}$ comes close to the black hole 
and becomes a part of it. 
Then, it starts to emit radiation.
As the particle loses energy, its wavelength increases. 
If the wavelength gets larger than the size of the black hole, 
then the particle can no longer stay in it. 
We can estimate the time scale of this process as $\sim a \log \f{a}{\sqrt{\s}}$, 
which is much shorter than that of the evaporation $\sim\f{a^3}{\s}$. 

Therefore, one of the important future works is 
to solve the wave equation in the self-consistent metric \Ref{eva_metric} more precisely
\footnote{See e.g. \ci{Akhmedov,Moskalets} for analysis of matter fields around the black hole.}. 
If we succeed in it, 
we should be able to understand 
how the information of the collapsing matter comes back 
and especially what happens to the baryon number conservation \ci{KY2}
\footnote{
There are many different approaches for the information problem. 
See e.g. \ci{Oda, Dundar, Oshita} for one on an infalling observer.}.

\section{Summary and discussion}\lb{s:Con}
Our solution tells what the black hole is. 
The collapsing matter becomes a dense object and evaporates eventually without forming a horizon or singularity. 
It has a surface instead of the horizon, but looks like an ordinary black hole from the outside. 
In the interior the non-trivial structure is formed, 
where the matter and the Hawking radiation can interact.
This can provide a possible solution to the information problem. 

There remain problems to be clarified in future. 
First, as we have mentioned, 
the important problem is to understand how the information comes back in this picture.
To do it, we need to solve the wave equation in the self-consistent metric \Ref{eva_metric}. 

Second, although we have assumed a constant $f$ to construct the metric \Ref{gen_metric}, 
we don't understand its meaning yet. 
In principle, 
$f$ should be determined by the dynamics of matters in the metric \Ref{gen_metric}. 
Therefore, it is interesting to evaluate $f$ concretely by considering a specific theory.

Third, the spherical symmetry has played the important role in our analysis.
In the real world, however, we need to consider a rotating black hole, 
the outside of which is described by the Kerr metric.
Although there is a conjecture on
the interior metric for a slowly rotating black hole \ci{KY2},
the general form is not known. 
It would be valuable if we can determine the interior metric by the 4D Weyl anomaly for the general case.

Fourth, we don't know yet how stable the metric \Ref{Tr} is 
for 
non-spherically symmetric perturbations. 
When investigating this problem, 
we need to be careful with the fact that the interior is not a fluid, as we have mentioned below \Ref{s_value}.

Finally, astrophysics has entered into a new stage 
by the launch of 
gravitational wave detectors. 
For a new physics of black holes it should be exciting to 
study an observable signal that exhibits some difference between 
the black holes in our picture and the conventional picture \ci{Dymnikova,Akhmedov2}.

\section*{Acknowledgment}
The authors thank the members of string theory group at National Taiwan University 
for valuable discussions. 
The present study was supported by
KAKENHI 16H07445 and the RIKEN iTHES project.
Y.Y. thanks Department of Physics in Kyoto University for hospitality.

\appendix 
\section{Derivation of \Ref{Vaidya_U2} and \Ref{Vaidya_V2}}\lb{ap:EMT_Vaidya}
We derive \Ref{Vaidya_U2} and \Ref{Vaidya_V2}. 
We first express the Vaidya metric \Ref{Vaidya} in the form of \Ref{UV}. 
We put $u=U$. Then, we introduce $V$ as a label of an ingoing null line following \Ref{r_u}: 
once an initial position for $r(U)$ in \Ref{r_u} is given, 
the solution is determined uniquely, which we denote by $\bar r (U,V)$. 
This plays roles of $r(U,V)$ in \Ref{UV}. 
Indeed, we have 
\begin{align}\lb{bar_r}
d\bar r &=\l(\f{\p \bar r}{\p U}\r)_V dU +\l(\f{\p \bar r}{\p V}\r)_U dV \nn \\
 &= -\f{\bar r- a}{2\bar r} dU +\l(\f{\p \bar r}{\p V}\r)_U dV,
\end{align}
replace $dr$ in \Ref{Vaidya} with this, and obtain 
\begin{equation}\lb{Vaidya2}
ds^2 = - 2 \l(\f{\p \bar r}{\p V}\r)_U  dUdV+\bar r(U,V)d\Omega^2,
\end{equation}
which means that $e^{\varphi(U,V)}=2 \l(\f{\p \bar r}{\p V}\r)_U$.

Under \Ref{T_UV}, 
we integrate \Ref{con_Ut} from $V_{out}$ to $V(>V_{out})$ along a fixed $U (\geq U_0)$: 
\begin{align}\lb{Vaidya_U}
(r^2 \bra T_{UU}\ket) -(r^2 \bra T_{UU}\ket)|_{V_{out}} 
 &= - \f{1}{2}\int^V_{V_{out}} dV' r \p_U r e^{\vp} \bra T^\mu{}_\mu \ket \nn\\
 &= - \int^V_{V_{out}} dV' \l(\f{\p \bar r}{\p V}\r)_U r \p_U r  \bra T^\mu{}_\mu \ket \nonumber \\
 &= - \int^{r(U,V)}_{r(U,V_{out}),U={\rm const.}} dr r \p_U r  \bra T^\mu{}_\mu \ket \nonumber \\
 &= \f{1}{2}\int^{r(U,V)}_{r(U,V_{out}),U={\rm const.}} dr (r-a(U))  \bra T^\mu{}_\mu \ket. 
\end{align}
Here, at the second line \Ref{Vaidya2} has been used; 
at the third line we have used the fact that $dr = dV \l(\f{\p \bar r}{\p V}\r)_U$ holds along a fixed $U$ (see \Ref{bar_r}); 
at the last line we employ \Ref{bar_r} again. 
Then, employing the boundary condition \Ref{bc_U}, we obtain \Ref{Vaidya_U2}. 

Next, we derive \Ref{Vaidya_V2}.
We integrate \Ref{con_Vt} with the assumption \Ref{T_UV} and the boundary condition \Ref{bc_V}: 
\begin{align*}
r^2 \TV &=(r^2 \TV)|_{U=-\infty} - \f{1}{2} \int^U_{-\infty}dU' r \p_Vr e^\vp \Tt \\
 &=- \int^U_{-\infty}dU' r (\p_Vr)^2 \Tt,
\end{align*}
where we have used $e^{\varphi(U,V)}=2 \l(\f{\p \bar r}{\p V}\r)_U$ in \Ref{Vaidya2}. 

Then, we estimate its order assuming that $a(U)$ varies slowly, 
$a(U)\sim $ const.
In this case, we can use \Ref{UV2} to have 
\begin{align*}
r^2 \TV &= - \int^r_{\infty,V={\rm const.}}dr' \f{1}{\p_U r} r' (\p_Vr)^2 \Tt\\
 &= \int^r_{\infty,V={\rm const.}}dr' r' \p_Vr\Tt\\
 &=\f{1}{2}\int^r_{\infty,V={\rm const.}}dr' (r'-a)\Tt.
\end{align*}
Using \Ref{Vaidya_p}, this becomes \Ref{Vaidya_VV}. 



\end{document}